\documentclass[aps,prl,twocolumn,superscriptaddress]{revtex4-1}

\usepackage{graphicx}
\usepackage{amsmath}
\usepackage{bm}
\usepackage{color}

\usepackage[bookmarks=true,colorlinks=true,urlcolor=blue,linkcolor=blue,citecolor=blue,breaklinks]{hyperref}

\usepackage{amssymb}
\usepackage{booktabs}

\begin{document}

\sloppy

\newcommand{\blue}[1]{{\textcolor{blue} {#1}}}
\newcommand{\red}[1]{{\textcolor{red} {#1}}}
\newcommand{\green}[1]{{\textcolor{green} {#1}}}
\title{Strain-driven chiral phonons in two-dimensional hexagonal materials}

\author{Habib Rostami}
\affiliation{Nordita, KTH Royal Institute of Technology and Stockholm University, Hannes Alfv\'ens v\"ag 12, 10691 Stockholm, Sweden}

\author{Francisco Guinea}
\affiliation{Imdea Nanoscience, Faraday 9, 28047 Madrid, Spain}
\affiliation{Donostia International Physics Center, Paseo Manuel de Lardiz\'abal 4, 20018 San Sebasti\'an, Spain}
\affiliation{ Ikerbasque. Basque Foundation for Science. 48009 Bilbao. Spain. }
\author{Emmanuele Cappelluti}
\affiliation{Istituto di Struttura della Materia-CNR (ISM-CNR), Trieste, Italy}


\begin{abstract}
Hexagonal two-dimensional materials with broken inversion symmetry (as BN or transition metal dichalcodenides)
are known to sustain chiral phonons with finite angular momentum, 
adding a further useful degree of freedom to the extraordinary entangled (electrical, optical, magnetic and mechanical) properties
of these compounds.
However,  because of lattice symmetry constraints, such chiral modes are constrained to the corners of the Brillouin zone,
allowing little freedom for manipulating the chiral features.
In this work, we show how the application of uniaxial strain leads to the existence
of new chiral modes in the vicinity of the zone center.
We also show that such strain-induced chiral modes, unlike the
ones pinned at the K points, can be efficiently manipulated by modifying the strain itself, which determines the position of these modes in the Brillouin Zone.
The results of the present paper add a new technique for the engineering of the quantum properties of two-dimensional lattices.
\end{abstract}

\maketitle

{\em Introduction--.} 
The possibility of exfoliating and/or growing two-dimensional materials
at the atomically-thin level
\cite{novoselov04,novoselov05,mak10,splendiani10,butler13,ajayan16,li17}
has paved the way to designing flexible systems
with a striking capability of converting mechanical deformations
into electronic or optical properties  \cite{vozmediano10,bertolazzi11,roldan15,amorim16}.
In the paradigmatic case of graphene, as well as for other systems with a hexagonal/triangular lattice,
like h-BN or transition-metal-dichalcogenides (TMDs)\cite{NoteTMDs},
the entanglement between electronic properties and the lattice is enriched by the
presence of two inequivalent sublattices. Such additional degree of freedom is conveniently cast in term
of a {\em spinor} vector \cite{castroneto09}  which, in the momentum space, is reflected in the occurrence of two inequivalently valleys at
the high-symmetry points K, K$^\prime$. The possibility of tuning the physical properties of two valleys,
and encoding there quantum information, has given rise to the concept of {\em valleytronics} \cite{tarasenko05,rycerz07}.
A peculiar feature of the graphene, BN and TMDs, due to their hexagonal lattice symmetry,
is that each valley is characterized by a remarkable chiral structure, with opposite chirality
of opposite K points.
Furthermore, the breaking of inversion symmetry
leads to a gap opening which is accompanied
by the onset of a finite Berry curvature  \cite{ando02,xiao_prl_07,xiaormp,ando15}.

Lattice modes (phonons) represent a further degree of freedom that can convey interesting quantum phenomena. Moreover, 
it was recently showed that also phonons in a hexagonal lattice with mass disproportion (e.g. BN or TMDs)
can carry  an angular momentum 
and sustain {\em chiral} modes, in the same way as electrons do.
\cite{zhang14,zhang15,zhu2018,chen2019,zhang20,komiyama21,chen21,ptok21,pirie22,notesquare}
Manipulating the properties of the lattice modes
can provide thus an alternative and/or complementary
scenario with promising perspectives for writing-in/reading-out quantum information.

The phonon dispersion of graphene presents a Dirac structure at the K, K$^\prime$ points in the subspace of the acoustic/optical modes.
In similar way as for the electronic Dirac states,
a key role is played by the breaking of inversion symmetry, driven by the mass disproportion,
and leading to the opening of a gap at K, K$^\prime$.
Just as for the electronic degrees of freedom, as pointed out in
the seminal work in Ref. \citenum{zhang15},
due to the crystal symmetry,
such chiral phonons
are pinned at the high-symmetry points K, K$^\prime$.
Symmetry arguments predict that chirality
concepts could apply as well at the
zone-center $\Gamma$ point, where
the degeneracy of the transverse and longitudinal
modes in the Cartesian basis can be
as well expressed in term of two (degenerate)
chiral modes with opposite sign.
Such opportunity is however not particularly useful
for practical purposes since the intrinsic degeneracy
of the two modes does not allow for a feasible
manipulation of the chiral degree of freedom.
Degeneracy splitting is thus a necessary requirement.
Within this perspective,
applying a real magnetic field, through
the electron-phonon coupling, has been shown to induce a chiral lattice polarization \cite{sonntag21},
whereas
uniaxial strain has been discussed to be detrimental
with respect to chirality since it favours linear
polarization \cite{huang09,mohiuddin09,wang13,doratotaj16}.
Finding the way of sustaining chiral modes in the closeness of the $\Gamma$ point is thus of the highest relevance since it could open the way for
a direct probe of chiral phonons by means
of optical means at ${\bf q}\approx 0$.

In this work we demonstrate that novel chiral phonons
can be conveniently generated and engineered 
close to the $\Gamma$ point
in the optical branches
of hexagonal systems with sublattice inequivalence (e.g. h-BN, TMDs, gapped graphene) by applying simple homogeneous uniaxial strain.
Furthermore, we show that such new chiral modes appear
in pairs of opposite chirality, 
and, unlike the chiral modes constrained
at the K, K$^\prime$ point,
the net momentum for each chirality
does not average to zero.
The closeness of the chiral phonons to the $\Gamma$ point
on the other hand opens the way for an optical probe.

\begin{figure}[!t]
\includegraphics[width=0.5\textwidth]{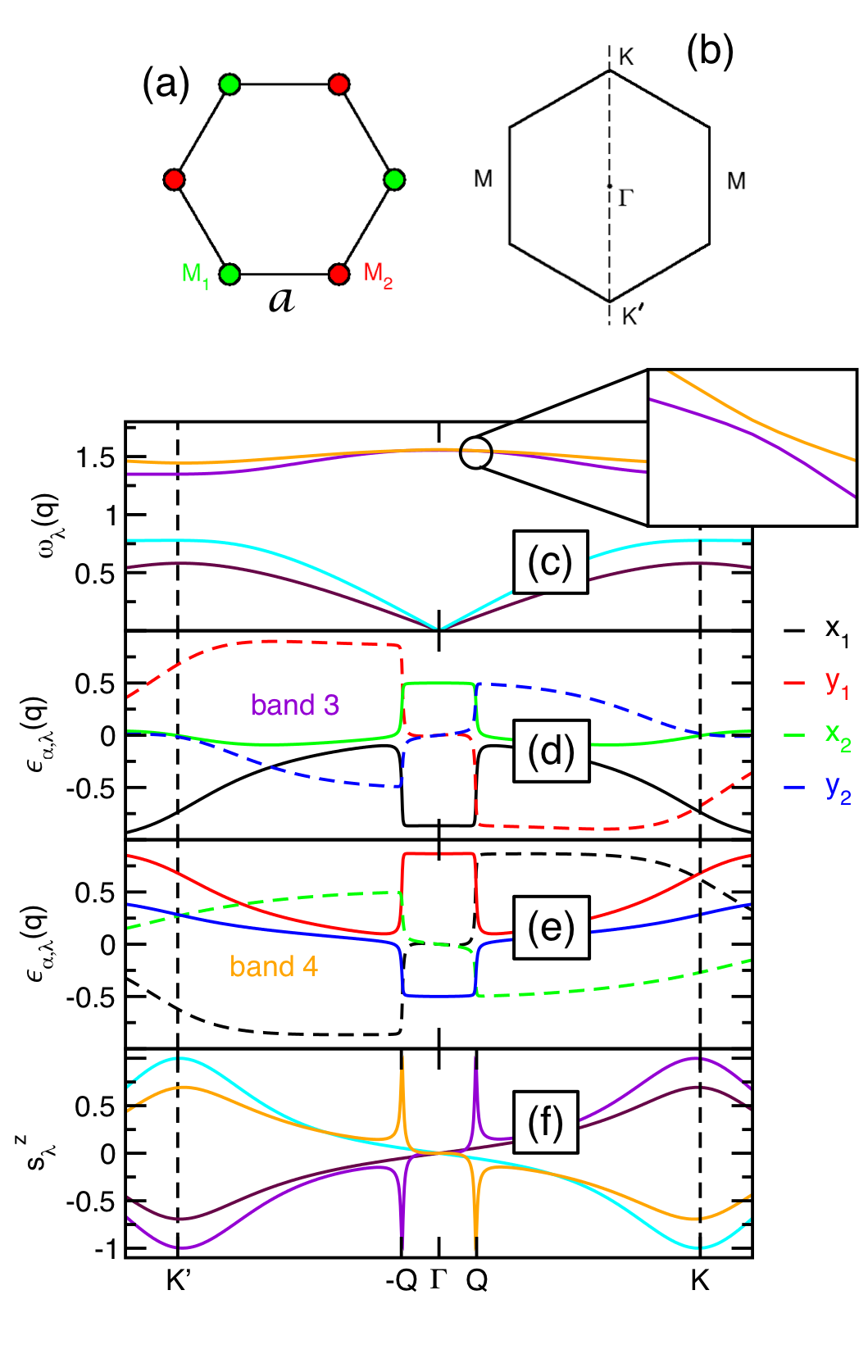}
\caption{(a) Lattice structure and (b) Brillouin zone
of a bipartite
hexagonal lattice. Atoms on the sublattice 1 have masses $M_1$,
atoms on the sublattice 2 have masses $M_2$.
(c) Phonon dispersion along the cut
K$^\prime$-$\Gamma$-K for the force-constant model
described in the text, upon an uniaxial tensile strain
$\varepsilon_+=$ 2 \%
along the $x$ direction.
Brown, cyan, violet and orange lines
denote phonon bands from $\alpha=1,\ldots,4$, respectively.
Inset: a magnification of the region close to the
avoided-band-crossing point Q. (d)-(e)
Corresponding eigenvectors of the lattice displacements of band 3 and 4.
Black, red, green and blue lines show the normalized lattice
displacement along $x_i$ and $y_i$ for atom $i$.
Solid lines represent the real component, whereas
dashed lines denote the imaginary part.
(f) Corresponding phonon circular polarization
$s_{z,\lambda}$ for all the four bands. Color code
as in panel (c).
}
\label{f-1}
\end{figure}

{\em Strain-driven chiral phonon modes--.} For a closer comparison with previous literature,
we begin considering the paradigmatic case
of a hexagonal lattice with inter-atomic distance $a$ and
two different masses $M_1$, $M_2$ in the two sublattices
(see Fig. \ref{f-1}a).
The corresponding Brillouin zone is also shown in Fig. \ref{f-1}b.
The elastic properties are described by means of a
nearest-neighbor
force-constant model with a radial spring constant $\phi_r$
and a transverse (in-plane) constant $\phi_t$.
Following a standard
derivation in literature (see Supplementary Material for further details \cite{SI}),
the lattice modes
are ruled by the dynamical matrix \cite{bornhuangbook}
which can be written as:
\begin{align}\label{eq:DSL}
\hat D({\bf q})=  
 \begin{pmatrix} {\hat G}/{M_1} & {\hat F_{\bf q}}/{\sqrt{M_1 M_2}} 
 \\[5pt]
 {\hat F^\ast_{\bf q}}/{\sqrt{M_1 M_2}} & 
{\hat G}/{M_2} 
\end{pmatrix} 
,
\end{align}
where $\hat G =\sum^3_{i=1} \hat {\Phi}_i$,
$F_{\bf q} =- \sum^3_{i=1} \hat {\Phi}_i e^{i {\bf q}\cdot{\bm \delta}_i}$,
and where
$\hat\Phi_1 = {\rm diag}[\phi_r,\phi_t]$ and $\hat \Phi_2 = \hat R^\dagger(\theta_0) \hat \Phi_1 \hat R(\theta_0)$ and $\hat \Phi_3 = \hat R^\dagger(-\theta_0)\hat \Phi_1 \hat R(-\theta_0)$.
Here $\hat R(\theta_0)$ represents the rotation matrix around the $z$-axes with $\theta_0=2\pi/3$.
In the following, in order to better show the novel results of our work,
we use the same force constant parameters employed in Ref. \citenum{zhang15},
$\phi_r=1$, $\phi_t=0.25$, with a atomic mass difference
$M_1=1$, $M_2=3$ that reproduces qualitatively the case of MoS$_2$.

The eigenvalues of the dynamical matrix provide the phonon dispersion $\omega_{\lambda}({\bf q})$
whereas the eigenvectors ${\bm \epsilon}_{\lambda}({\bf q})$ rule the lattice displacements and
carry the information about the chiral character.
 This can be captured by switching
the  representation
of the lattice displacements
from the Cartesian basis
$|x_1\rangle$, $|y_1\rangle$, $|x_2\rangle$, $|y_2\rangle$,
to the chiral basis defined by the
$|R_\alpha\rangle=
\left(|x_\alpha\rangle+i|y_\alpha\rangle\right)/\sqrt{2}$,  $|L_\alpha\rangle=
\left(|x_\alpha\rangle-i|y_\alpha\rangle\right)/\sqrt{2}$,
($\alpha=1,2$).
It is thus convenient to introduce
the phonon circular polarization \cite{zhang15,SI},
defined as
\begin{align}
 s_{z,\lambda}({\bf q})=
\sum_{\alpha=1,2}\left(
|\epsilon_{R,\alpha,\lambda}({\bf q})|^2-|\epsilon_{L,\alpha,\lambda}({\bf q})|^2\right) 
,
\end{align}
where $\epsilon_{R/L,\alpha,\lambda}$ are the phonon eigenvectors of the dynamical matrix expressed in the chiral basis.
The phonon circular polarization is strictly related
to the phonon orbital angular momentum,
which can be defined for each phonon band 
as $L_{z,\lambda}({\bf q}) = \hbar s_{z,\lambda}({\bf q}) \{n_{\rm B}[\omega_{\lambda} ({\bf q})/T]+1/2\}$,
where $n_{\rm B}[x]=1/[\exp(x)-1]$ if the Bose-Einstein factor.

The phonon dispersion displays four bands
(we number them $\lambda=1,\ldots,4$, from lower to higher energies)
with two optical branches
degenerate at $\Gamma$ with $E_{2g}$ symmetry and
two acoustic branches.
In agreement with Ref. \citenum{zhang15}, for the perfect hexagonal lattice, we find chiral modes with opposite chirality at the
K, K$^\prime$ points for bands 2 and 3.
An inspection of the eigenvectors show that such chiral modes
have a pure-sublattice character, with the mode of band 2 (3) involving
chiral lattice displacements only of the heavier (lighter) sublattice.

We will show that new physics and new chiral modes appear
upon applying an anisotropic strain,
described by the strain tensor
$\varepsilon_{ij}=(\partial_i u_j+\partial_j u_i)/2$, where $\partial_i u_j$ is derivative of the {\em strain} lattice displacement $u_j$
along the direction $i$ \cite{timo}.
On general ground, the above tensor
can be viewed as a combination, an isotropic biaxial average strain $\varepsilon_0=(\varepsilon_{xx}+\varepsilon_{yy})/2$,
of a tensile ($\varepsilon_+>0$) and a compressive ($\varepsilon_->0$)
orthogonal components,
aligned along the angles \cite{atan}
\begin{eqnarray}
\theta_+ = \theta_-
+\frac{\pi}{2}
&=&
\frac{1}{2} \arctan \left( \frac{2\varepsilon_{xy}}{\varepsilon_{xx}-\varepsilon_{yy}} \right)
,
\end{eqnarray}
and with respective strain values $\varepsilon_\pm = \varepsilon_0 \pm \Delta \varepsilon/2,$
where 
\begin{eqnarray}
\Delta \varepsilon
&=&
\sqrt{
\left(\varepsilon_{xx}-\varepsilon_{yy}\right)^2
+
4\varepsilon_{xy}^2
}.
\end{eqnarray}
Within the framework of a force-constant analysis,
anisotropic in-plane strain affects the phonon properties
through two main effects: ($i$) the elongation/compression of the bonds,
tuning the spring constants $\phi_r(R)$, $\phi_t(R)$ whose values depend intrinsically
on the interatomic distance $R$;-- similar to the case of hopping energies in electronic tight-binding models \cite{Kane_1997, ando_prb_2002, Sasaki_2005, Vozmediano_2010, Mariani_prb_2012, rostami_prb_2013, rostami_prb_2015} ($ii$) the geometrical change of the angles between neighbor atoms, implying different mixing of the $x$ and $y$ components of the lattice displacements.
For the sake of completeness, we retain both effects, although each of them, separately, is sufficient to generate new
chiral modes. The mathematical implementation of anisotropic strain within the force-constant model 
follows a straightforward procedure and we refer to the Supplementary Material for technical details \cite{SI}.

In Fig. \ref{f-1}c we show the phonon dispersion
along the vertical cut K$^\prime$-$\Gamma$-K
(see panel b) for an uniaxial tensile strain
$\varepsilon_+=$ 2 \% ($\varepsilon_-=0$ ) along
the $x$ (armchair) direction ($\theta_+=0$).
We label the bands as $\lambda=1,\ldots,4$, from lower to higher energies.
At the $\Gamma$ point (${\bf q}=0$)
we observe the well-known splitting of the $E_{2g}$ phonons
in two non-degenerate modes $\omega_{\Gamma^-}$, $\omega_{\Gamma^+}$
($\omega_{\Gamma^-} < \omega_{\Gamma^+}$)
with eigenvectors aligned along $\theta_+$, $\theta_-$.
As a general rule
the softer mode $\omega_{\Gamma^-}$ 
corresponds to lattice displacements
along the tensile axis $\theta_+$ of strain direction (in this case the $x$ direction), while
the harder mode $\omega_{\Gamma^+}$ is associated with the compressive strain
along $\theta_-$. Such splitting, for the tensile strain along $x$, implies
a ``band crossing" at momenta $\pm Q$
along the $y$-axis K$^\prime$-$\Gamma$-K cut based on the simple model here considered with transverse-optical (TO) modes
harder than the longitudinal-optical (LO) ones \cite{NLOTO}.

\begin{figure}[t]
\includegraphics[width=0.46\textwidth]{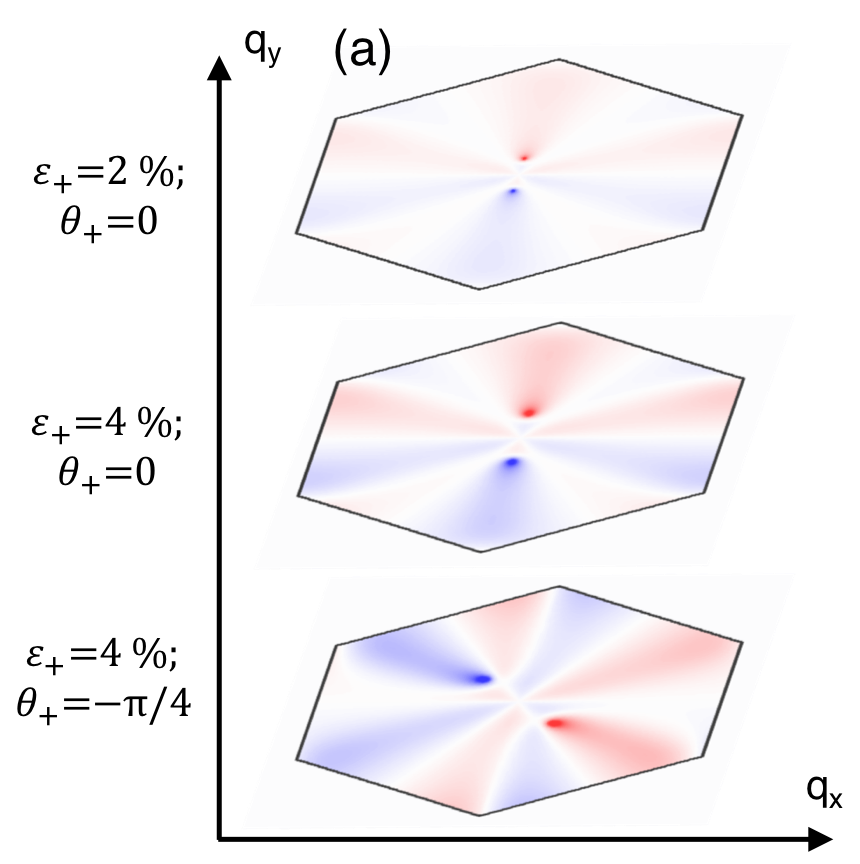}
\includegraphics[width=0.38\textwidth]{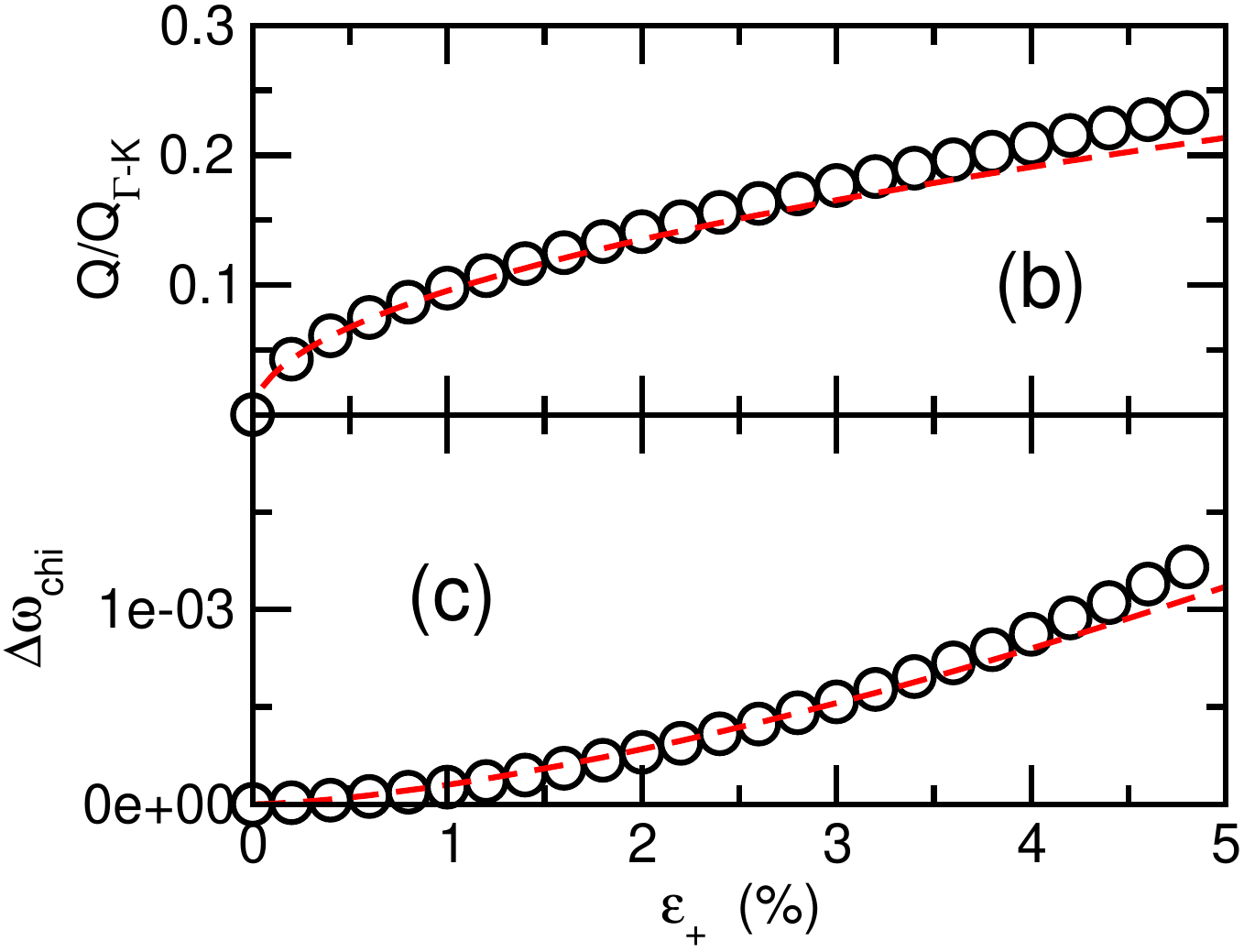}
\caption{(a) Color map of $\Delta s_{z,\lambda}$ for the optical phonon band 4 upon different
values and directions of applied strain. Red and blue regions represent positive and negative phonon circular polarizations, respectively.
Panels (b)-(c): dependence of crossing point ${\bf Q}$ and of the phonon gap $\Delta\omega_{\rm chi}$
on the strain magnitude for the top case shown in panel (a). Black symbols represent results using
the full force-constant model; the dashed red line using the two-band model.
}
\label{f-tune}
\end{figure}

A closer look at the phonon dispersion (see inset of Fig. \ref{f-1}c)
shows however that such ``phonon-band-crossing"
is however just apparent but it is actually
an ``avoided-band-crossing" with a finite gap $\Delta\omega_{\rm chi}$
between the two phonon branches.
A deeper understanding of the physics here at work
can be attained from the analysis of the phonon mode
eigenvectors $\epsilon_{\alpha,\lambda}({\bf q})$ (Fig. \ref{f-1}d,e).
As expected, close to the $\Gamma$ point,
in the strained case, band 3 and band 4 
are characterized by lattice displacements
almost purely aligned along $x$ and $y$ respectively,
with atom 2 moving counter-phase versus atom 1.
The opposite occurs of course for momenta larger
than the ``crossing'' point, where the order of the bands
is reversed.
A true band-crossing would happen if
the eigenvectors would be purely real.
However, at finite momenta, a small imaginary component
is always unavoidably present.
The presence of such finite imaginary part
prevents a true crossing, expected between bands with orthogonal lattice displacements,
and the switch between the lattice displacements of band 3 for
$|q_y| \le Q$ and band 4 for $|q_y| \ge Q$
occurs through a sudden {\em chiral twist}
where the real $x$ component acquires an imaginary $y$ component
and vice versa.
Right at the momentum of the avoided-crossing point,
${\bf q}=\pm Q$, the eigenmodes of band 3 and 4 are described
exactly by the chiral basis. This is remarkably shown in Fig. \ref{f-1}f where we
illustrate the phonon circular polarization for each phonon band.
Here we can clearly see that, besides the well-known chiral phonons
at K, K$^\prime$, in bands 2 and 3, {\em new} chiral phonons appear upon anisotropic strain in bands 3 and 4 close to the $\Gamma$ point,
with opposite circular phonon polarization at $\pm Q$.
It is worth to stress here that such novel chiral phonons
present striking different properties with respect to
the standard chiral phonon pointed out in Refs. \cite{zhang15,chen2019}.
Indeed, while chiral phonons discussed in Refs. \cite{zhang15,chen2019}
are locked at the high-symmetry points K, K$^\prime$ at the edge of the Brillouin zone, and they obey a threefold symmetry, the chiral phonons arising
upon anisotropic strain have a {\em tunable} location in the Brillouin zone
controlled by the strength and by the direction of the strain.
Since for vanishing strain $|\varepsilon| \to 0$ the  ${\bf q}$-momentum
of such chiral phonon approaches zero, $|{\bf Q}| \to 0$,
for small strain these phonons are expected to appear close to the $\Gamma$ point, 
where spectroscopic optical probes can be effective.

{\em Low-energy description--.} In order to highlight the role of the anisotropic strain,
we analyze for each phonon band the change in the phonon circular polarization
induced by the anisotropic strain, namely
$\Delta s_{z,\lambda}=s_{z,\lambda}(\hat{\varepsilon})-s_{z,\lambda}(\hat{\varepsilon}=0)$.
The tunability of such new chiral phonons by means of  strain 
is shown in Fig. \ref{f-tune}a where we plot in a color map
$\Delta s_{z,\lambda}$ of the optical phonon band $\lambda=4$
for different magnitudes and directions of the applied strain.
We see that increasing strain shifts the momentum $\pm {\bf Q}$
of chiral phonons at larger values, whereas the presence of a shear component
rotates ${\bf Q}$ away from the crystallographic axes.
As a general rule, the couple of such strain-induced chiral phonons, with opposite
circular phonon polarization, appears along the direction in the reciprocal space
{\em perpendicular} to the strain tensile component in real space.

The dependence of ${\bf Q}$ on the strain magnitude
is shown in Fig. \ref{f-tune}b, while in
panel c we plot the value of the phonon gap $\Delta \omega_{\rm chi}$
between band 4 and 3.
A full control and prediction of the properties
of such new chiral phonons upon strain can be analytically achieved
by introducing an effective reduced model.
To this aim we 
employ a Schrieffer-Wolf transformation
restricted to the relevant bands 3 and 4,
which are described at the $\Gamma$ point
by counter phase displacements
(along $x$ and $y$) of atoms 1 and 2 with
a weight dictated by the masses:
$\epsilon_3(\Gamma)=(1/\sqrt{M_1},0,-1/\sqrt{M_2},0)$,
$\epsilon_4(\Gamma)=(0,1/\sqrt{M_1},0,-1/\sqrt{M_2})$.
Technical details are reported in the Supplementary Material \cite{SI}.
In such $2 \times 2$ Hilbert space it is convenient
to work directly in the chiral basis,
$v_L=(v_3+iv_4)/\sqrt{2}$,
$v_R=(v_3-iv_4)/\sqrt{2}$.
The resulting phonon properties at the
linear order in strain are described thus
by the dynamical matrix
$\hat{D}_0({\bf q})+D_\varepsilon({\bf q})$.
were $\hat{D}_0({\bf q})$ accounts for the lattice properties
in the absence of strain, while
$D_\varepsilon({\bf q})$ 
contains the corrections due to an anisotropic strain.
More explicitly,
close to the $\Gamma$ point, we have
\begin{eqnarray}
\hat{D}_0({\bf q})
&=&\omega^2_\Gamma
\begin{pmatrix}
1-a_0|{\bf q}|^2 &
a_1 q_L^2 \\
a_1 q_R^2 & 
1-a_0|{\bf q}|^2
\end{pmatrix}, 
\label{eqH0}
\end{eqnarray}
where 
$q_{L/R}=q_x\pm iq_y$,
and where $a_0=0.077$, $a_1=0.028$
are the relevant parameters of such ${\bf k} \cdot {\bf p}$-like
expansion for the phonon bands. The numerical value
of $a_0$, $a_1$ is dictated by the microscopical parameters
$\phi_r$, $\phi_t$, $M_1$, $M_2$ and their analytical
expression is provided in the Supplementary Material \cite{SI}.

The resulting phonon bands in unstrained case
are described thus by
$\omega_{\pm}({\bf q}) = \gamma_\pm({\bf q})\omega_\Gamma$
where $\gamma^2_{\pm}({\bf q})=1-(a_0\pm a_1)|{\bf q}|^2$. 
The strain correction term on the other hand
can be computed at the leading order in the
${\bf q} \to 0$ limit and it
reads:
\begin{eqnarray}
\hat{D}_\varepsilon({\bf q}=0)
&=&\omega^2_\Gamma
\begin{pmatrix} 
-2\alpha_0 \varepsilon_0 &
\alpha_1 \varepsilon_R
\\
\alpha_1 \varepsilon_L
&
-2\alpha_0 \varepsilon_0
\end{pmatrix},
\label{Hstrainq0}
\end{eqnarray}
where
$\varepsilon_{L/R}=\varepsilon_{xx}-\varepsilon_{yy}
\pm i2\varepsilon_{xy}$, $\alpha_0=1.5$ and $\alpha_1=0.15$.
The phonon bands computed by diagonalizing $\hat{D}_0({\bf q})+D_\varepsilon(0)$
read thus $\omega_{\pm}({\bf q}) = \gamma_{\pm,\varepsilon}({\bf q}) \omega_\Gamma$, where
\begin{align}
&\gamma_{\pm,\varepsilon}^2({\bf q})
=
1-a_0|{\bf q}|^2
-
2\alpha_0 \varepsilon_0
\nonumber\\
&
\pm\sqrt{
\left[
a_1(q_x^2-q_y^2)
-\alpha_1 
\left(
\varepsilon_{xx}-\varepsilon_{yy}
\right)
\right]^2
+
4\left[
a_1q_xq_y
+
\alpha_1 \varepsilon_{xy}
\right]^2
}
.
\label{wpm}
\end{align}
Note that,
although the parameters $a_i$, $\alpha_i$ can be as well easily evaluated within
the minimal nearest-neighbor force-constant model (see Supplementary Material),
the validity of Eqs. (\ref{eqH0})-(\ref{wpm})
holds true in full generality for realistic
phonon bands where the parameters $a_i$, $\alpha_i$
can be extracted from first-principle calculations and/or from
direct experimental probes. Within the present framework,
limited to a quadratic expansion in ${\bf q}^2$, we can now analytically identify the momentum $Q$ where a band anticrossing occurs and chiral phonons appears. For the teachful example of uniaxial tensile strain along the $x$ direction, $\varepsilon_+\neq 0$, $\varepsilon_-=0$, $\theta_+=0$, we find $Q_x=0$, $Q_y=\pm\sqrt{\alpha_1\varepsilon_+/a_1}$, in perfect agreement with the numerical results for the full phonon band dispersion (Fig. \ref{f-1}c). For generic case, including shear strain, we find that the momentum ${\bf Q}_\pm = \pm Q(\cos\phi,\sin\phi)$
is univocally determined by the strain tensor, with the angle $\phi$ aligned along the compression direction, $\phi=\theta_-$, and $Q^2$ being determined by the strain difference between the two main components:
\begin{align}
 Q^2 = \left|\frac{\alpha_1}{a_1}\right| \Delta\varepsilon. 
\end{align}
It is worth to mentioning that, in addition to the above analysis,
using in Eq. (\ref{eqH0})-(\ref{Hstrainq0})
an expansion for small strains hence small ${\bf q}$'s,
is very efficient in determining the location of the
new chiral modes, but it also predicts a {\em true} phonon band crossing,
with two degenerate modes at ${\bf Q}$, not a well-defined
chiral character. A true gap, and thus a robust chiral character, is
recovered however in a more realistic modelling
when a higher order dependence on
the momenta is retained. From a dimensional analysis,
since $|Q|$ scales with the square root of
the strain amplitude, one can realize that
the next leading order effects are driven by the cubic ${\bf q}^3$
terms in Eq. (\ref{eqH0}), which lead to a correction
$\hat{D}_0({\bf q})\to 
\hat{D}_0({\bf q})+\hat{D}_1({\bf q})$
where
$\hat{D}_1({\bf q}) \propto A |{\bf q}|^3\hat{\sigma}_3$.
The gap 
$\Delta\omega_{\rm chi}$
between the two new chiral modes appearing
under strain is predicted thus to behave as
$\Delta\omega_{\rm chi} \propto (\Delta\varepsilon)^{3/2} \sin(3\phi)$. 
The strain dependence of the gap $\Delta\omega_{\rm chi}=\omega_+({\bf Q})-\omega_-({\bf Q})$ between the two chiral modes is also shown in Fig. \ref{f-tune}c by comparing the numerical results
obtained by the effective phonon dispersion evaluated with the force-constant model
with the effective two-band model. Once again we find a striking agreement.

It should be stressed the role of the 
gap $\Delta\omega_{\rm chi}$.
Indeed, like for the chiral modes at the K points,
it is the presence of such gap that makes
the chiral modes physically observable.
Assuming for characteristic two-dimensional
systems (BN, TMDs) that the frequency of the optical modes lies
in the range $\omega_\Gamma \approx 300-1400$ cm$^{-1}$ \cite{molina15,zhangph15,geick66,reich05,sanchezportal12,cai17},
and considering realistic values of the strain, the gap can be
of the order of a fraction or one cm$^{-1}$,
beyond the possibility of a spectroscopy probe.
Alternatively, the effect of such strain-induced
chiral modes might be detected in macroscopic quantities,
through the different populations
$n_{\rm B}[\omega_+({\bf Q})/T]\neq
n_{\rm B}[\omega_-({\bf Q})/T]$,
where $T$ is the temperature.
It should be noticed indeed that, 
due to such new chiral modes, 
the system acquires a finite
``chiral wavevector'' defined as
$\bar{\bf Q}
=
\sum_{\lambda,\alpha_\pm}
{\bf Q}_\alpha s_{z,\lambda}({\bf Q}_\alpha)
n_{\rm B}[\omega_\lambda({\bf Q}_\alpha)]$.
This is significantly different from
the chiral modes at the K points in the absence of strain
where the threefold symmetry enforces $\bar{\bf Q}=0$.
The temperature, which has here a clear relevant role
ruling the finite different population, should be compared
with the optical phonon frequencies $\omega_\Gamma\approx \omega_+({\bf Q}),\omega_-({\bf Q})$.
At room temperature such ratio is still quite small,
preventing probably a direct experimental probe of $\bar{\bf Q}$.
Interesting perspectives could be prompted however from time-resolved
pump-probe approaches. A common trait of two-dimensional hexagonal materials
is indeed that pump-induced energy, initially stored in the electronic
degrees of freedom, can be quickly transfer to few lattice modes,
corresponding to intravalley and intervalley scattering.
Those modes (including ${\bf Q}_\alpha$)
can thus get {\em hot}, with effective temperatures
that can reach the order to $10^3$ K, making the possibility
of probing the effects of a finite $\bar{\bf Q}$ more accessible.
Further promising scenarios can come from near-field optical techniques,
where finite ${\bf q}$ optical excitations can be launched,
and couple to the chiral phonons considered here.
Tuning the near-field conditions to better probe ${\bf q}\approx {\bf Q}_\alpha$,
with a circular polarized light, might be efficiently result
in selectively exciting phonon modes with a specific chirality.

A final mention concerning the possible ways
of probing the new (strain-driven) chiral phonons at small ${\bf q}$'s
is through their effect on the acoustic modes.
As discussed above, at ${\bf Q}_\alpha$
the physics of the avoided crossing point
is strictly reflected in the optical branches,
inducing modes with opposite chiralities $s_{z,3/4}({\bf Q}_\alpha)\approx \pm 1$.
However, a small but sizable chirality content
is present at ${\bf Q}_\alpha$ also in the acoustic branches,
with $s_{z,1}({\bf Q}_\alpha)\approx -0.16$
and $s_{z,2}({\bf Q}_\alpha)\approx 0.15$,
with a net $s_{z,1}({\bf Q}_\alpha)+s_{z,2}({\bf Q}_\alpha)\approx -0.01$.
Although clearly less strong than in the optical branches,
it should keep in mind that such difference applies to modes
that are {\em strongly} non degenerate.
Furthermore, the low-energy of the acoustic lattice modes
makes them easily thermally populated and highly relevant
for the transport properties.
The strain-induced chirality might be thus
feasibly probed through the macroscopical effects
related to the small but finite chiral content of the acoustic modes.

{\em Acknowledgements.} This work was supported by Nordita and the Swedish Research Council (VR 2018-04252). F. G. acknowledges funding  from the European Commision, under the Graphene Flagship, Core 3, grant number 881603, and by the grants NMAT2D (Comunidad de Madrid, Spain),  SprQuMat and SEV-2016-0686, (Ministerio de Ciencia e Innovación, Spain).

\bibliography{bibliography_chiral_phonon}

\begin{thebibliography}{52}
\bibitem [{Not()}]{NoteTMDs}%
  \BibitemOpen
  \href@noop {} {}\bibinfo {note} {The crystal structure of single-layer $MX_2$
  TMDs can be viewed from the top as an hexagonal lattice, i.e. a triangular
  lattice with a proper basis defined by $M$ on the sublattice A and the
  top/bottom $X$ atoms on the sublattice B.}\BibitemShut {Stop}%
\bibitem [{ata()}]{atan}%
  \BibitemOpen
  \href@noop {} {}\bibinfo {note} {{This expression is technically valid only
  for $\varepsilon_{xx}>\varepsilon_{yy}$. For generic strain one should
  replace $\arctan({y/x})$ with ${\rm arctan2}(x,y) =
  -i\ln[(x+iy)/\sqrt{x^2+y^2}]$.}}\BibitemShut {Stop}%
\end{thebibliography}

\end{document}